\documentclass[american]{IEEEtran}
\usepackage[T1]{fontenc}
\usepackage[latin9]{inputenc}
\usepackage{textcomp}
\usepackage{amsmath}
\usepackage{graphicx}
\usepackage{amssymb}

\makeatletter

\providecommand{\tabularnewline}{\\}

\makeatother

\makeatother

\usepackage{babel}

\makeatother

\usepackage{babel}

\begin{document}

\title{Generating Scale-free Networks with Adjustable Clustering Coefficient
Via Random Walks}

\author{\IEEEauthorblockN{Carlos~Herrera and Pedro~J.~Zufiria} \\
 \IEEEauthorblockA{Depto. Matem\'atica Aplicada a las Tecnolog\'ias
de la Informaci\'on \\
 ETSI Telecomunicaci\'on, Universidad Polit\'ecnica de Madrid \\
 Ciudad Universitaria s/n, 28040 Madrid, Spain\\
 Email: carlos@hyague.es, pedro.zufiria@upm.es}}

\maketitle

\begin{abstract}
This paper presents an algorithm for generating scale-free networks
with adjustable clustering coefficient. The algorithm is based on
a random walk procedure combined with a triangle generation scheme
which takes into account genetic factors; this way, preferential attachment
and clustering control are implemented using only local information.
Simulations are presented which support the validity of the scheme,
characterizing its tuning capabilities. 
\end{abstract}

\section{Introduction}

Network modelling has become a very active research field after the
discovery that many different complex systems share some essential
common features which can be gathered in a network model \cite{dorogovstev2003evolution}.
Although network nodes and links can represent very different entities
depending on the phenomenon being analyzed, still some common characteristics
seem to be ubiquitous in many models. For instance, common patterns
usually appear in social networks (\cite{barabasi2002evolution,onnela2007structure,liljeros2001web}),
biology networks (\cite{white1986structure,jeong2000large}), technological
networks (\cite{faloutsos1999power,chen2002origin}) or information
networks (\cite{barabasi1999emergence,redner1998popular})%
\footnote{This classification of networks according to its nature was proposed
by Newman in \cite{newman2003structure} where a larger number of
references for each type of network can be found.%
}. Many of these features are non-trivial so that the traditional Erdös-Rènyi
(ER) model \cite{erdos1960evolution} for random graphs is not sufficient
to explain the behavior of these systems.

The first feature appearing in real networks which was not gathered
by the ER model was the small world effect, defined by two factors: slow
increase of network diameter with network growth and the existence
of a unexpectedly high number of triangles in the network (clustering).
In order to mimic these properties, Watts and Strogatz proposed a
new model in \cite{watts1998collective}. Nevertheless, this model
could not represent an additional property also found in many real
networks: the distribution of the number of neighbors (degree distribution)
follows a power-law, which is very different from the distributions
predicted by early models (e.g., exponential distribution in the ER
model).

In \cite{barabasi1999emergence} the Barabási-Albert (BA) network
generation model was presented where network growing nature and preferential
attachment were proved to be two essential features for obtaining
scale-free networks which follow a power-law in the degree distribution.
The preferential attachment stands for the fact that new vertices
added to the network are attached preferentially to high-degree vertices.
In case that the preference is linear, the probability to get connected
to a given vertex is proportional to its degree. The BA model implements
this preferential attachment using global network information to compute
such probability: \begin{equation}
p_{i}=\frac{k_{i}}{\sum_{j=1}^{n}k_{j}},\ n=\text{total number of nodes}.\label{eq_prob_pref}\end{equation}

The existence of a scale-free structure in many real networks has
motivated the appearance of a number of new network models trying
to reproduce at least one of the already mentioned three main characteristics
of real networks (clustering, long tail degree distribution, short
diameter)%
\footnote{Alternatively, some models try to mimic other network properties.
For example, the goal of the model presented in \cite{karrer2010random}
is to ensure that the network shows an arbitrary subgraph distribution.
In \cite{toivonen2006model}, the property to be reproduced is the
existence of communities like the ones observed in real social networks;
this could be considered as a generalization of clustering control.%
}. Different approaches have been used: some models are based on a
static network size \cite{newman2001random,davidsen2002emergence},
while others work on growing networks \cite{barabasi1999emergence,holme2002growing,vazquez2003growing}.

In general, it is expected that the process of adding a new vertex
in real world networks would not require the availability of such
global information. Along this line, several authors have studied
alternative local schemes (employing rules that only involve a vertex
and its neighbors) to generate scale-free networks without the use
of global parameters \cite{li2010emergence,smith2006realistic,vazquez2003growing,kumar2000web,krapivsky2001organization,dorogovtsev2000structure,jost2002evolving}.

Among them, the use of random-walkers to select node attachment in
a network-growth algorithm has been suggested in \cite{albert2002statistical}
and successfully employed in \cite{vazquez2000knowing,saramaki2004scale,evans2005scale}.
In general, the use of the proposed schemes has been justified on
the assumption that a random walk of arbitrary length $l$ will end
up on a vertex $i$ of degree $k_{i}$ with probability given in equation
(\ref{eq_prob_pref}), i.e., random walks are assumed to generate
a pure preferential selection procedure (to be used as the basis of
a preferential attachment scheme). The analytical characterization
of these random walk models has been performed under some mean-field
hypotheses, so that preferential attachment is studied but no other
network features are considered. In \cite{vazquez2003growing} the
correlation between clustering and degree is analysed also under the
mean-field hypotheses, but the tuning of the clustering coefficient
is not addressed.

In this paper, alternative random walk selection schemes are presented
which allow for the control of both preferential attachment and clustering
coefficient in the process of growing a scale-free network. The schemes
depend on the transition probability distribution of the random walk
in a manner that each path sample may have a different size due to
a genetic factor. It is shown that the appropriate selection of the
mentioned transition probability distribution allows for the tuning
of the clustering coefficient of the generated network.

The paper is organized as follows. Section \ref{sec:mod} starts
presenting a motivation and the goals of the proposed model for network
generation; then such model is described in detail. In Section \ref{sec_simu}
simulation results supporting the validity of the model are presented.
Finally, Concluding Remarks and Future Work lines are summarized in
Section \ref{sec_conc}.

\section{Model\label{sec:mod}}

\subsection{Motivation and goals}

Many real world networks are very complex systems governed by several
fundamental characteristics. So far, existing network models can only
gather some of these features, which may or not be sufficient for
the aim of the analysis. Hence, the construction of new more elaborated
models addressing the emergence and behavior of additional characteristics
is a relevant challenge. As mentioned in the Introduction, one big
step in terms of explaining complex networks was the BA Model, where
the emergence of a power-law in the degree distribution was explained
via two simple assumptions: growing and preferential attachment (PA).
The BA model has been a fundamental reference although it presents
some limitations.

On the one hand, several authors \cite{evans2005scale,saramaki2004scale}
have pointed out the difficulties of a practical implementation of
preferential attachment policies: as defined in the BA model, when
a new node is about to join the network, it requires to know the degree
of all nodes in the whole network in order to calculate the probability
of linking to each existing node. This scheme does not seem to successfully
explain the behavior of real-life stages, such as a blogger linking
to a web page or a person making new friends (obviously, they do not
have or do not employ global network structure information). A new
model based only on local schemes was presented in \cite{saramaki2004scale},
suggesting that PA can be obtained from a random walk (at least in
an approximate manner). In \cite{evans2005scale} this random walk
based model is generalized so that the degree distribution can have
an exponent different from $\gamma=3$ if a certain fraction $p_{v}$
of edges is created purely at random (i.e, the new node is linked
with a randomly chosen existing node, without implementing a walk).

On the other hand, there is also one important feature which cannot
be taken into account when employing the original BA model. Although
this model performs better than Erdös-Rényi (ER) model concerning
the degree distribution, it cannot produce the high clustering coefficient
which has been observed in many real networks%
\footnote{Although there are no analytical results for the clustering coefficient
in the BA model, it is known (\cite{albert2002statistical,newman2003structure})
that it decays with network size $C\sim N^{-0.75}$ while in real
networks $C$ is independent from $N$. %
} (see table \ref{tab:Coeficientes-de-clusterin}, presenting results
from \cite{mislove2007measurement} for online social networks). The
clustering coefficient of real networks is known to be higher than
the one provided by a purely random model, and its value depends on
the nature of the network. It usually takes high values for social
networks (for example $C=0.79$ for imdb actor network \cite{watts1998collective})
but there are some networks which exhibit a power-law with a much
lower clustering level ($C=0.011$ for a P2P network \cite{ripeanu2002mapping}).
In \cite{holme2002growing} a mechanism for triangle formation was
proposed which allowed the control of the clustering coefficient;
such mechanism is furtherly developed and employed in this paper.

\begin{table}
\caption{\label{tab:Coeficientes-de-clusterin}Clustering coefficient in real
networks, and in models of the networks of the same size}

\noindent \centering{}\begin{tabular}{|c|c|c|c|}
\hline 
Network  & $C$  & $C/C_{ER}$  & $C/C_{BA}$\tabularnewline
\hline
\hline 
Flickr  & 0.313  & 47.2  & 25.2\tabularnewline
\hline 
LiveJournal  & 0.330  & 119.0  & 17.8\tabularnewline
\hline 
Orkut  & 0.171  & 7.24  & 5.27\tabularnewline
\hline 
Youtube  & 0.136  & 36.9  & 69.4\tabularnewline
\hline
\end{tabular}
\end{table}

In \cite{fowler2009model} some characteristics in the social networks
which are related to genetic factors are presented. Concretely, it
is shown that the clustering coefficient is one of those heritable
network metrics. In fact, there are people who are very likely to
introduce friends to any other friends, whereas some other people
prefer to keep their friends apart from each other.

In this paper, a new network growing scheme is presented where a triangle
formation scheme is furtherly developed to include a genetic factor
as the basis for a clustering control mechanism. This genetic factor
in the nodes (known to happen in real networks) combines in a very
adequate manner with a random walk based node selection procedure,
so that only local information is employed in the whole edge addition
process.

\subsection{Model description}

As mentioned before, the main aim of the model proposed in this paper
is to generate scale-free networks whose clustering coefficient can
be controlled by using only local information. The scale-free network
is generated via a growing scheme which employs random walks as a
local approximation to the preferential attachment criterion. 
The model proposed here is grounded on a modification of the model
presented by Evans and Saramäki (ES model) \cite{saramaki2004scale}.
The ES model is defined as follows: 
\begin{itemize}
\item Initial condition: start with a network of $n_{0}$ vertices. 
\item Growth: each time step a vertex and $m$ edges are added to the network.
Note that $m\leq n_{0}$. 
\item Linking by a random walk (RW): the new vertex $v_{new}$ is joined
with $m$ existing vertices which are selected the following way:
a random existing vertex $v_{s}$ is chosen, then a $l$-step random
walk starting from $v_{s}$ is performed; the arrival vertex $v_{e}$
obtained at the end of the walk is linked to $v_{new}$. 
\end{itemize}
Our model makes use of some random walk properties which happen to
be very useful to control the appearance of triangles in the network.
First of all, let us consider $v_{new}$ has been linked to a first
selected network node $v_{s}$; obviously this link does not generate
any triangle. Now it can be seen that a $l=1$ walk starting from
$v_{s}$ will provide a new node that, if also linked to $v_{new}$,
will generate a triangle. This way, selecting nodes via successive
$l=1$ walks would add $m-1$ triangles to the network. This fact
suggests the possibility of implementing a triangle generation control
scheme based on selecting new starting random points for the next walk
(avoiding triangle formation), proportionally combined with successive
$l=1$ walks (forcing triangle formation). This approach was the first to be
analysed: such control mechanism does not behave accurately when reproducing
low levels of clustering coefficient. In fact, for a given clustering
control parameter, a significant variance was observed, violating our
design principle of fine clustering coefficient control (numerical
results of this behavior will be showed in section \ref{sec_simu}).

\begin{figure}
\begin{centering}
\includegraphics[width=1\columnwidth]{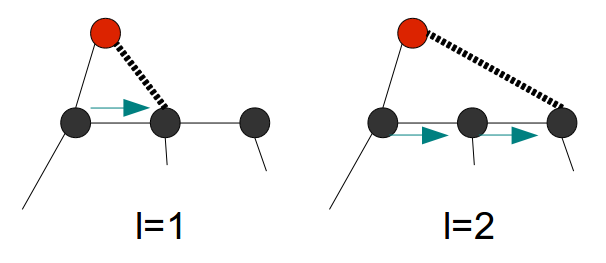} 
\par\end{centering}

\caption{\label{fig:clustering-control}The clustering coefficient is controlled
by the following mechanism: if a $l=1$ walk is carried out from the last joined
vertex $v_{s}$, we ensure that a triangle is formed. However, if $l=2$
we only generate a triangle if $v_{s}$ and $v_{e}$ are already connected.}

\end{figure}

Alternatively, random walks with $l=2$ were employed. Note that if
$l=2$ is chosen almost no triangle will be added to the network.
In fact, if the seed network with $n_{0}$ vertices does not have
any triad, it is straightforward to demonstrate that no triangle will
be added (the probability of $v_{s}$ and $v_{e}$ being already connected
will be zero). The implications of using $l=1$ or $l=2$ are illustrated
in Figure \ref{fig:clustering-control}.

Therefore, controlling the number of $l=1$ and $l=2$ random walks
allows for an accurate tuning of the number of triangles added to
the network. This might be implemented by just assigning $p_{1}$
probability to $l=1$ walks and $1-p_{1}$ for the $l=2$ case. However,
we propose to include this control in a node attribute which is assigned
at node's {}``birth'', inspired by the already mentioned results
in \cite{fowler2009model}. In our proposal, any vertex $p$ is assigned
a probability $p(v_{i})$ which reflects the genetic factor mentioned
there upon a certain distribution $f(p)$. This probability remains
constant during vertex life and will determine the length of the random
walks starting from such node (i.e., a 1-step walk happens with probability
$p(v_{s})$, being $v_{s}$ the last linked vertex; a 2-step walk
is selected otherwise).

Following the explained principles our algorithm is defined as follows: 
\begin{enumerate}
\item Start with a network of $n_{0}$ vertices. Each vertex is assigned
an attribute $p(v_{i}),\ i=1\ldots n_{0}$ according to the random distribution
$f(p)$. 
\item A vertex $v_{s}$ is chosen randomly. 
\item A random walk $l>1$ is performed from $v_{s}$, randomly choosing
at each step a neighbor of the current vertex. The arrival vertex
$v_{e}$ is marked. 
\item Start a new walk from the last marked vertex $v_{l}$. With probability
$p(v_{l})$ this will be 1-step walk; otherwise will be a 2-step walk.
Mark the arrival vertex. 
\item Repeat step 4 $m-1$ times. Note that $m\leq n_{0}$. 
\item Add one vertex to the network. Add $m$ links between the new vertex
and the $m$ marked nodes. Assign to the new vertex a probability
$p(v_{i})$ according to $f(p)$ %
\footnote{The reason why any chosen vertex is first marked instead of directly
linked to the new vertex (in step 3 or 4) is because it is desirable
for the network to remain unchanged during the addition of the whole
$m$ edges. This procedure is common in many implementations of scale-free
networks models such as BA, since the hypothesis of the network remaining
unchanged during the vertex addition process is used in the mean-field
equations model which supports that preferential attachment produces
power-laws in degree distribution \cite{newman2003structure}. In
random walk based models, this unchanged network hypothesis is even
more important, since the addition of an edge to the last visited
vertex can severely change the trajectory of the following random
walk.%
}. 
\item Repeat steps 2 to 6 ($n-n_{0}$ times). 
\end{enumerate}
Hence, the algorithm has the following design parameters: number of
nodes $n,$ number edges to be attached per vertex $m$, and a probability
distribution $f(p).$ Note that, as it is usual in growing network
models, $m$ allows to control the average degree since $\left\langle k\right\rangle =2m$.
Concerning the distribution $f(p),$ in order to simplify the interpretation
of results, a binomial distribution has been chosen to illustrate
the scheme in this paper; so there is a fraction $cc$ of nodes with
$p(v_{i})=1$ having the rest of the nodes $p(v_{i})=0$. It is expected
that different distributions may lead to different community structures
of the resulting networks. 

Another interesting issue is the selection of $l>1$ for the first
walk, since it is different from previous random walk based models.
Although in \cite{saramaki2004scale} is stated (and supported by
numerical simulations) that a walk of length 1 should be enough to
produce a valid preferential attachment, we have found problems with
some networks, where there is still a significant correlation between
the neighbors average degree and how frequently a certain vertex is
marked by a 1-step walk (in pure preferential attachment, the vertex
selection criterion is not biased by the vertex neighbors degrees).

\section{Simulation Results\label{sec_simu}}

A number of simulations have been carried out to check the performance
of the proposed model along its two main goals: generating a power-law
in the degree distribution and controlling the clustering coefficient%
\footnote{In our simulations $c_{i}=\begin{cases}
\frac{2e_{ij}}{k_{i}(k_{i}-1)} & k_{i}> 1\\
0 & k_{i}\leq 1 \end{cases}$, and $C=\frac{1}{N}\sum_{i}c_{i}$. There is also a different clustering
degree definition for the global graph in \cite{barrat2000properties}
but it is not used in this paper. %
}.

\begin{figure}
\begin{centering}
\includegraphics[width=0.6\columnwidth]{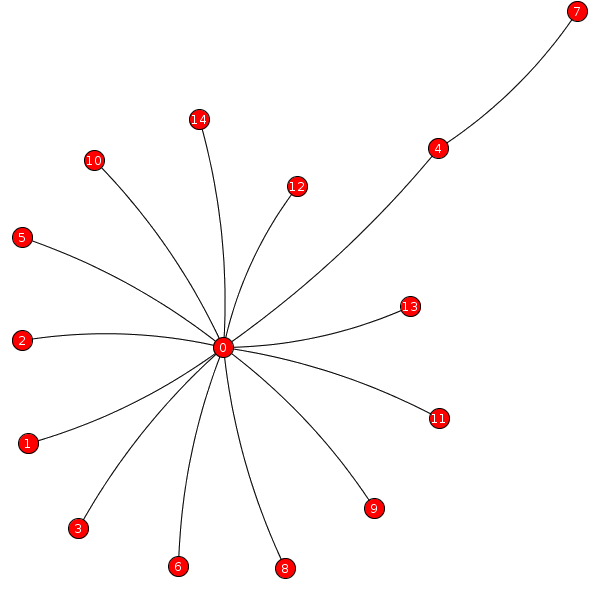} 
\par\end{centering}

\caption{\label{fig:win-takes-all}An inappropriate election of $n_{0}$ may
drive to a {}``winner-takes-all'' effect. In this simulation $m=1$
and $n_{0}=2$, produce a star-like graph where a power-law degree
distribution cannot emerge by a growing process based on preferential
attachment.}

\end{figure}

\begin{figure}
\begin{centering}
\includegraphics[width=0.8\columnwidth]{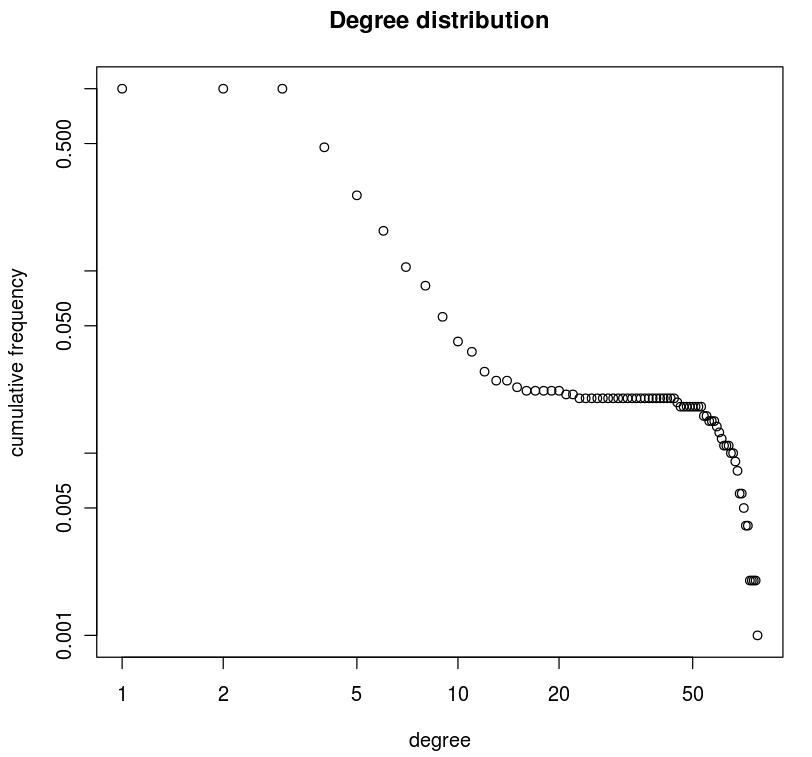} 
\par\end{centering}

\caption{\label{fig:step-power}Simulation with $N=10^{3}$, $m=2$, using
a full connected graph of $n_{0}=20$ nodes as a seed. It clearly
produces a {}``step-like'' deviation from a power-law tail in the
cumulative degree distribution.}

\end{figure}

First of all, random walk models require the construction of an initial
seed connected network with $n_{0}$ vertices; it is important to
point out the influence of the topology of such initial network on
the final outcome. As it also happens in the BA model, a non-zero
probability must be assigned to the initial isolated vertices (in
most implementations this is done by setting a certain parameter $a$
in the distribution so that $p(k)\sim k+a$). In order not to bias
the first walk, a regular lattice must be chosen where the degree
$k_{0}$ is equal for all $n_{0}$ nodes (this is equivalent to the
mentioned $a$ parameter in the BA model). Besides, even if we start
with a network accomplishing those requirements, the performance of
the model might be affected by the values selected for $n_{0}$ and
$k_{0}$. Precisely, if $n_{0}$ is too small, there is a high probability
of a so called {}``winner-takes-all'' phenomenon to happen as it
can be seen in Figure \ref{fig:win-takes-all}. On the other hand,
choosing a large value for $k_{0}$ might also create some undesired
effects. In this case, choosing a fully connected graph (i.e $k_{0}=n_{0}-1$)
might produce a deviation from the power-law behavior in the degree
distribution as can be seen in Figure \ref{fig:step-power}. To overcome
this situation, a ring graph has been designed, with $n_{0}=\max(10,m)$,
which seems to behave properly in most of the stages; hence, all the
simulation results presented below make use of this seed. In addition,
as pointed out in the model description, a length $l=7$ for the first
walk per added node has been used, in order to avoid the dependence
on the neighbors degree during the selection process.

\subsection{Scale-free emergence}

\begin{figure}
\begin{centering}
\includegraphics[width=0.8\columnwidth]{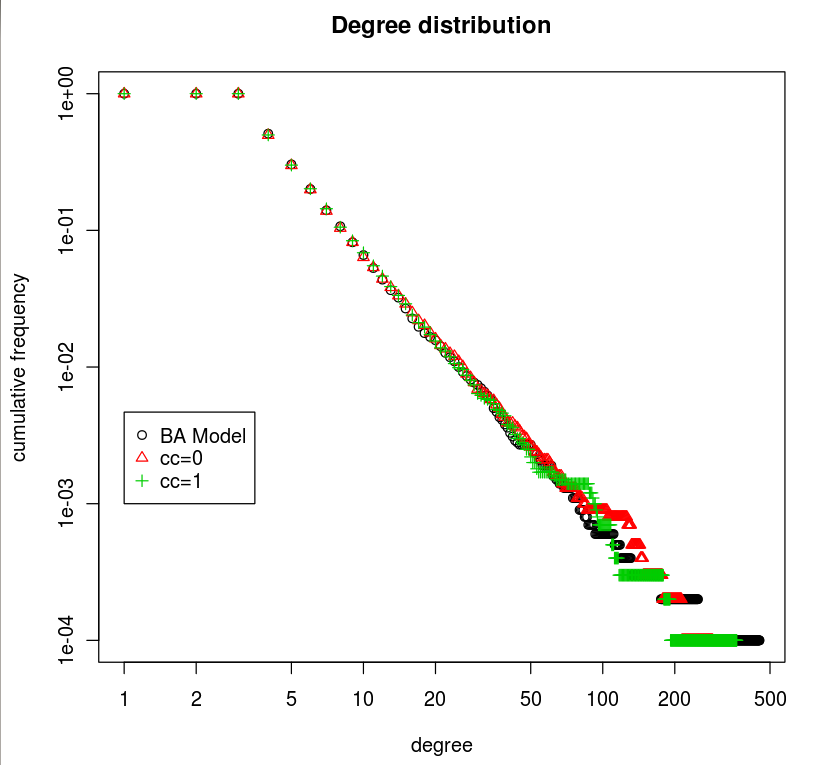} \label{fig_power} 
\par\end{centering}

\caption{\label{fig:power-law-cc} A power-law is obtained in the degree distribution,
very close to BA pure preferential attachment model. These results
persist under any change in the clustering control parameter $cc$.}

\end{figure}

Once a proper selection procedure of the initial seed has been settled,
we now focus on the first design criterion: the emergence of a scale-free
network during the growing process. In Figure \ref{fig:power-law-cc},
it can be seen that a power-law is produced (original BA model%
\footnote{For the BA model $a=2$ was selected so that it did agree with our
model where a ring (i.e $k_{0}=2$) is used as a seed. %
} simulation is also included, in order to allow comparison). In addition,
it is shown that this behavior does not depend on the value assigned
to the clustering control parameter $cc$, as it can be seen in the
representation of results for the two extreme values of this parameter.
This power-law regime is independent from $cc$ as well as from the
size of the network $n$ and the average degree $\left\langle k\right\rangle =2m$;
hence, the model proposed here behaves like other preferential attachment
based models \cite{barabasi1999emergence,evans2005scale}.

\subsection{Clustering coefficient control}

\begin{figure}
\begin{centering}
\includegraphics[width=0.8\columnwidth]{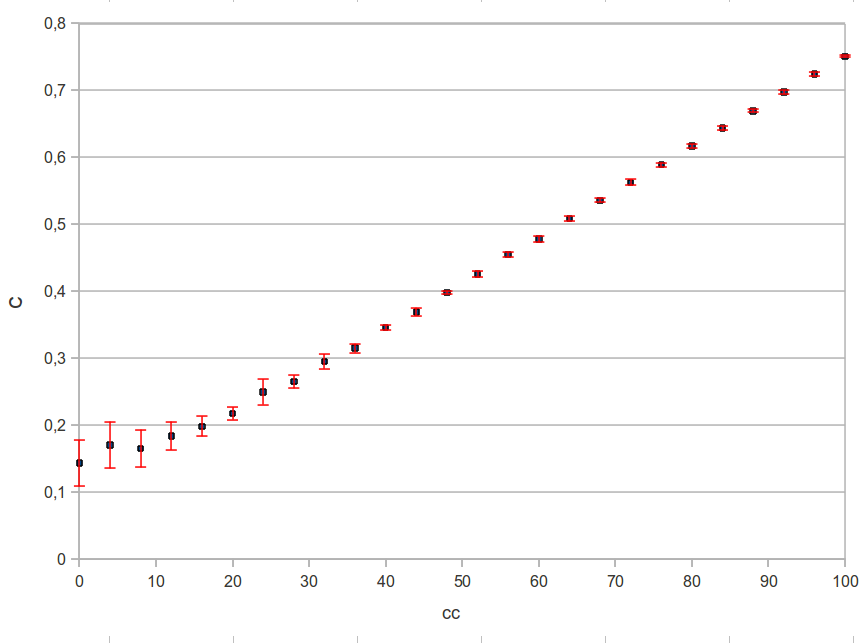} 
\par\end{centering}

\caption{\label{fig:modelA} Clustering coefficient control by the new random
starting point strategy. The model shows a significant level of variance
(red bars represent standard deviation for 20 runs) for small values
of $cc$.}

\end{figure}

\begin{figure}
\begin{centering}
\includegraphics[width=0.8\columnwidth]{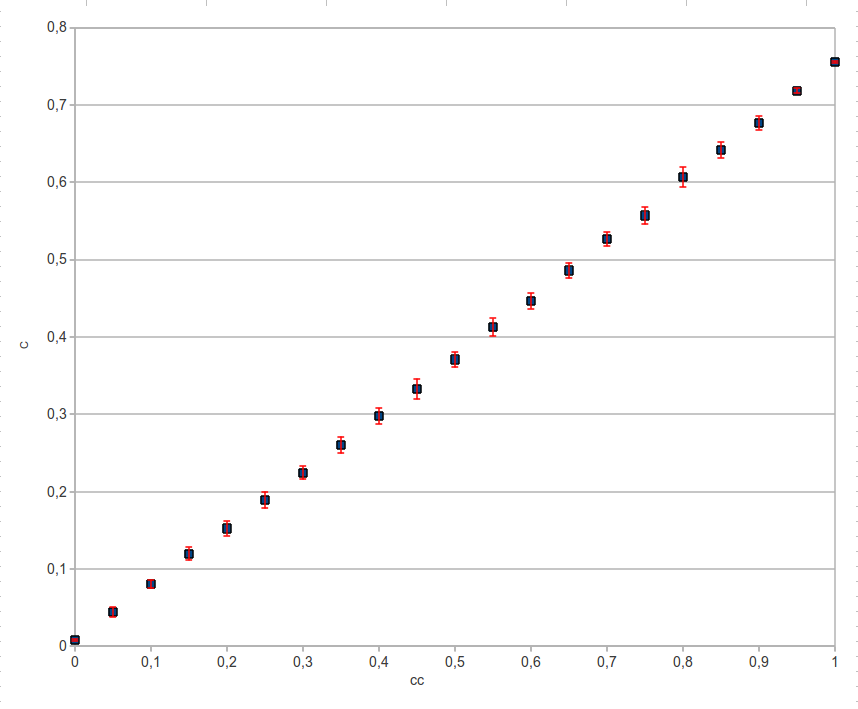} 
\par\end{centering}

\caption{\label{fig:modelB} Clustering coefficient control by the 2-step walk
strategy. In this case, the model shows a much better performance
following a linear relationship characterized by $C=0.74986\cdot cc$ with
$R\text{\texttwosuperior}=0.99965167$.}

\end{figure}

As mentioned earlier, the control of the clustering coefficient is
performed by changing the value of the probability $cc$ that a node
is assigned {}``1'' length value in the binomial distribution characterizing
genetic factors. We start by presenting the simulation results which
correspond to the first (more intuitive) approach mentioned in Section
\ref{sec:mod}; such approach suggested the use of a new random starting
vertex for avoiding triangle formation. The results show that this
approach drives to a high level of variance for small values of $cc$
as it can be seen in Figure \ref{fig:modelA}. On the other hand,
the second approach based on a 2-step mechanism produces a much better
performance as shown in Figure \ref{fig:modelB}. Both figures show
the clustering coefficient dependence on networks with $N=10^{4}$
nodes and $\left\langle k\right\rangle =4$. Mean clustering coefficient
(blue points) and standard deviation (red bars), obtained for twenty
runs for each value of $cc$, are also presented.

\begin{figure}
\begin{centering}
\includegraphics[width=0.8\columnwidth]{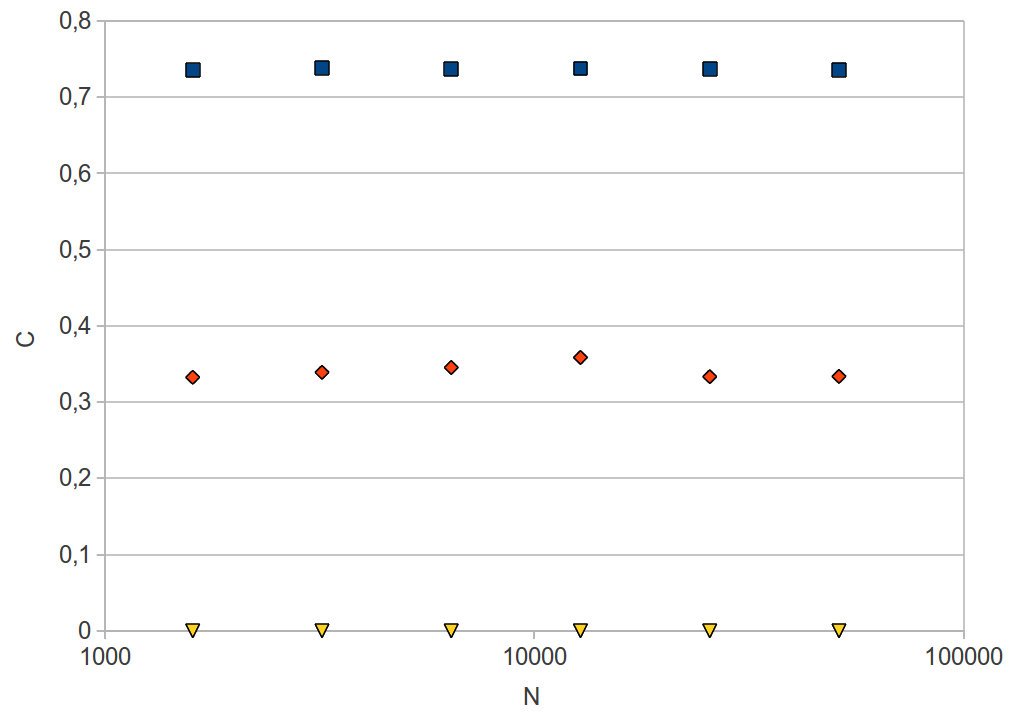} 
\par\end{centering}

\caption{\label{fig:cc-N} Clustering coefficient variation with network size
$N$, for fixed $m=2$. Results are presented for $cc=0$ (yellow),
$cc=0.5$ (orange) and $cc=1$ (blue).}

\end{figure}

Two additional tests were performed for the proposed model regarding
its control capability of the clustering coefficient. The first test
proves that the clustering coefficient remains constant for a given
$cc$ if the network keeps growing. This result is supported by Figure
\ref{fig:cc-N}, where the degree is log-plot and the results from
$N=1600$ to $N=50\cdot10^{3}$ are presented.

\begin{figure}
\begin{centering}
\includegraphics[width=0.8\columnwidth]{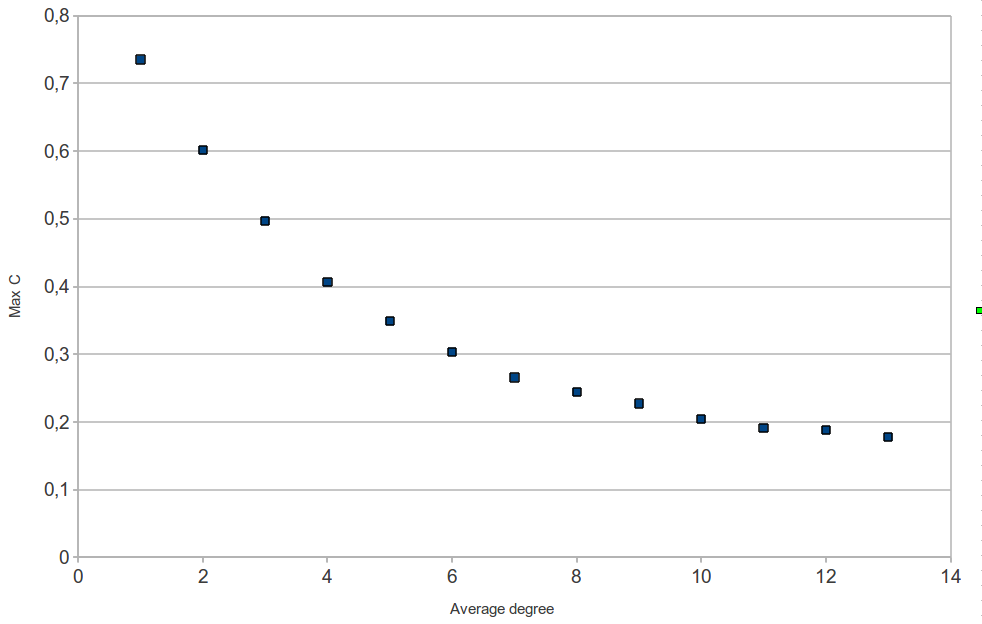} 
\par\end{centering}

\caption{\label{fig:cc-n} Clustering coefficient decay with average degree.}

\end{figure}

\begin{figure}
\begin{centering}
\includegraphics[width=0.8\columnwidth]{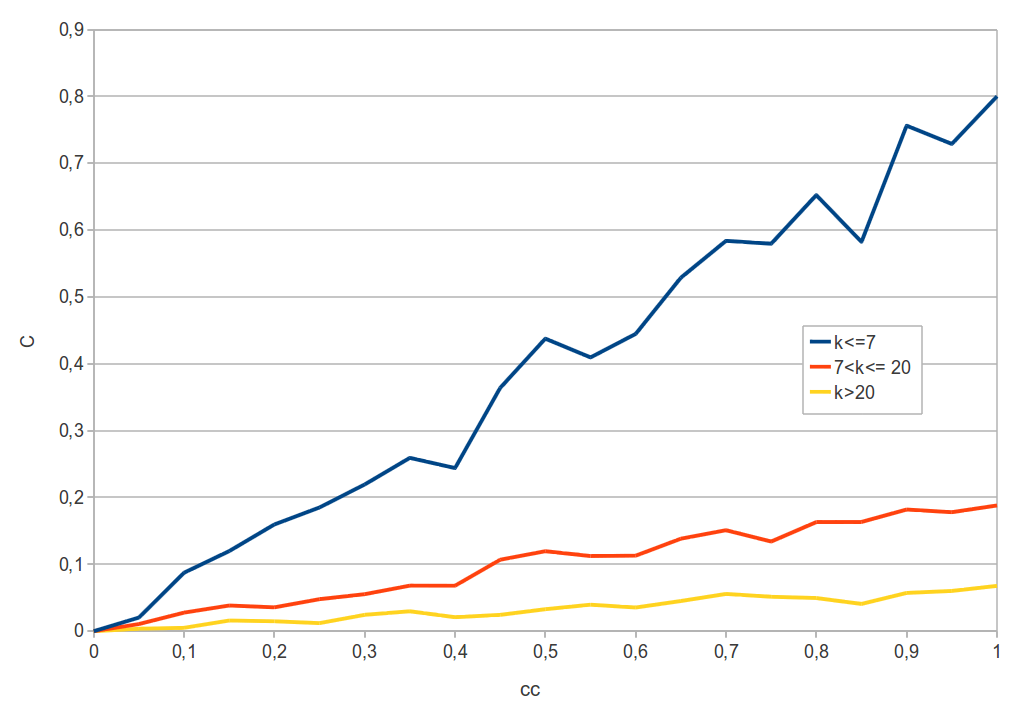} 
\par\end{centering}

\caption{\label{fig:cc-control-degree-bias} Clustering coefficient control
problem for high degree nodes (same simulation parameters as in Figure
\ref{fig:modelB}). Nodes are divided 
into 3 groups according to their degree: the clustering
control coefficient $cc$ has smaller influence on the most connected nodes.}

\end{figure}

The last simulation performed was intended to study the variation
of the maximum clustering coefficient that can be generated by the
model, that is $cc_{\max}$, as $m$ is increased. The results presented
in Figure \ref{fig:cc-n} show a decay of the clustering coefficient
as $m$ increases; this expected result shows a behavior equivalent
to other tunable clustering network models \cite{holme2002growing,serrano2005tuning}.
In fact, some authors have proposed the use of an alternative clustering
coefficient definition since large values of $m$ do bias the degree-clustering
correlation in scale-free networks when the standard definition of
clustering coefficient is employed. The problem can be summarized
as follows: very high-degree nodes (so called {}``hubs'') have very
few chances of having a high clustering coefficient; this is due to
the fact that it would require most of their neighbors to be connected
among themselves, producing a full graph around the hub, which does
not fit in a scale-free structure. To support this intuition, a simulation
has been performed, whose results are presented in Figure
\ref{fig:cc-control-degree-bias}, showing that 
the hubs do not reach high values of clustering coefficient.
A new formulation for the clustering coefficient to avoid this degree
bias has been proposed in \cite{soffer2005network}. The performance
of our model on this new definition has not been analysed yet.

\section{Conclusions and future work}

\label{sec_conc}

In this paper, a new scheme for generating scale-free networks with
power law degree distribution and tunable clustering coefficient has
been presented. The scheme is grounded on a combination of random
walk and triangle generation procedures together with a genetic factor
implementation. These elements allow for an accurate tuning of the
clustering coefficient, making use of local information exclusively.
As a consequence, this proposed scheme seems to explain the generation
of real networks in a more realistic manner. The presented simulations
support the validity of the scheme, characterizing its tuning capabilities.

Further research is being carried out in several directions. On the
one hand, the sensitivity of preferential attachment policies to the
random walk length $l$ is being analysed. On the other hand, some
work is also being developed in reproducing additional network metrics
by using different $f(p)$ distributions for vertex characterization.
An appropriate selection of $f(p)$ can potentially drive to a network
where not only average clustering coefficient is controlled, but also
the whole clustering coefficient distribution over the network. The
distribution $f(p)$ could also be made to depend on some other network
metrics (e.g., the degree, so that $f(p,k)$) in order to reproduce
some correlations between network metrics observed in real networks.
Finally, it is worth mentioning that generalizations of this model,
based on a network growth driven exclusively by local interaction
and intrinsic network attributes, can be implemented in different
ways. For instance, 
some variants proposed in previous random walk models \cite{evans2005scale}
can be easily incorporated to the network model presented here.



\section*{Acknowledgements}

The authors want to acknowledge the financial support of Orange Spain,
in the framework of C\'atedra Orange at the ETSI Telecomunicaci\'on in the 
Universidad Polit\'ecnica de Madrid (UPM). The work has
been also partially supported by projects MTM2010-15102 of Mi\-nis\-terio
de Ciencia e Innovaci\'on, and Q10 0930-144 of the UPM, Spain.

\bibliographystyle{plain} 
\bibliography{article-models-final-rev-ok}

\end{document}